\documentclass[sigconf, nonacm]{acmart}

\usepackage{amsmath,amsfonts}
\usepackage{algorithm, algpseudocode}
\usepackage{graphicx}
\usepackage{textcomp}
\usepackage[usenames,dvipsnames]{xcolor}
\usepackage{xspace}
\usepackage{fontawesome}
\usepackage{multirow}
\usepackage{rotating}
\usepackage{tikz}
\usepackage{soul}
\usepackage{makecell}
\usepackage{url}
\usepackage{enumitem}
\usepackage{booktabs}
\usepackage{fontawesome}
\usepackage{pifont}
\usepackage{tcolorbox}
\usepackage[caption=false]{subfig}
\usepackage[export]{adjustbox}
\usepackage{tabularx}
\usepackage{array}
\usepackage{colortbl}
\usepackage{relsize}
\usepackage{listings}

\tcbuselibrary{skins}
\newcolumntype{Y}{>{\centering\arraybackslash}X}
\newcolumntype{L}{>{\raggedleft\arraybackslash}X}

\setcopyright{none}
\renewcommand\footnotetextcopyrightpermission[1]{}

\newcommand{\verylightgray}[1]{\cellcolor{gray!10}{#1}}
\newcommand{\lightgray}[1]{\cellcolor{gray!22}{#1}}
\newcommand{\gray}[1]{\cellcolor{gray!33}{#1}}
\newcommand{\darkgray}[1]{\cellcolor{gray!45}{#1}}
\tcbset{
  tab2/.style=
  {
    enhanced,
    fonttitle=\bfseries,fontupper=\normalsize\sffamily,
    colback=white, 
    colframe=black,
    colbacktitle=darkGray,
    coltitle=white,
    center title
  }
}

\newcommand{\lstbg}[3][0pt]{{\fboxsep#1\colorbox{#2}{\strut #3}}}
\lstdefinelanguage{diff}{
  basicstyle=\ttfamily\scriptsize,
  morecomment=[f][\lstbg{red!20}]-,
  morecomment=[f][\lstbg{green!20}]+,
  morecomment=[f][\textit]{@@},
}

\newcommand{\ie}{\emph{i.e.,}\xspace}
\newcommand{\eg}{\emph{e.g.,}\xspace}

\newcommand{\etal}{\emph{et~al.}\xspace}
\newcommand{\secref}[1]{Section~\ref{#1}\xspace}

\newcommand{\tabref}[1]{Table~\ref{#1}\xspace}

\newboolean{showcomments}
\setboolean{showcomments}{true}

\ifthenelse{\boolean{showcomments}}
  {\newcommand{\nb}[2]{
    \fbox{\bfseries\sffamily\scriptsize#1}
    {\sf\small$\blacktriangleright$\textit{#2}$\blacktriangleleft$}
   }
  }
  {\newcommand{\nb}[2]{}
  }

\begin{document}
\title{Evaluating the Impact of Post-Training Quantization on Large Language Models for Code Generation}

\author{Alessandro Giagnorio}
\affiliation{%
  \institution{Universit\`a della Svizzera Italiana}
  \country{Switzerland}
}
\email{alessandro.giagnorio@usi.ch}

\author{Antonio Mastropaolo}
\affiliation{%
	\institution{William \& Mary}
	\country{United States}
}
\email{amastropaolo@wm.edu}

\author{Saima Afrin}
\affiliation{%
	\institution{William \& Mary}
	\country{United States}
}
\email{safrin@wm.edu}

\author{Massimiliano Di Penta}
\affiliation{%
	\institution{University of Sannio}
	\country{Italy}
}
\email{dipenta@unisannio.it}

\author{Gabriele Bavota}
\affiliation{%
	\institution{Universit\`a della Svizzera Italiana}
	\country{Switzerland}
}
\email{gabriele.bavota@usi.ch}

\begin{abstract}
Large Language Models (LLMs) have shown an impressive capability in code generation. The LLM effectiveness generally increases with its size: The higher the number of LLM's trainable parameters the better its ability to implement code. However, when it comes to deploying LLM-based code generators, larger LLMs pose significant challenges related to their memory (and, consequently, carbon) footprint. A previous work by Wei \etal proposed to leverage quantization techniques to reduce the memory footprint of LLM-based code generators  without substantially degrading their effectiveness. In short, they studied LLMs featuring up to 16B parameters, quantizing their precision from floating point 32 bits down to int 8 bits and showing their limited impact on code generation performance. Given the fast pace at which   LLM capabilities and quantization techniques are evolving, in this work we present a differentiated replication of the work by Wei \etal in which we consider (i) on the one side, more recent and larger code-related LLMs, of up to 34B parameters; (ii) the latest advancements in model quantization techniques, which allow pushing the compression to the extreme quantization level of 2 bits per model parameter and; (iii) different types of calibration datasets to guide the quantization process, including code-specific ones. Our empirical evaluation reveals that the new frontier for LLM quantization is 4-bit precision, resulting in an average memory footprint reduction of 70\% compared to the original model without observing any significant decrease in performance. Additionally, when the quantization becomes even more extreme (3 and 2 bits), a code-specific calibration dataset helps to limit the loss of performance. 
\end{abstract}


\maketitle

\section{Introduction} \label{sec:introduction}

Large Language Models (LLMs) opened up new frontiers in code generation. However, their usage also introduced new concerns. Developers could leverage closed models, often available through APIs and often also integrated into the development environments. This is the case, for example, of GitHub CoPilot \cite{copilot} or GPT \cite{gpt4}. While in such cases, the training and inference are performed on third-party servers, using such models introduces data privacy issues \cite{abs-2401-07348,wu2024unveiling,huang2023security}. Also, such models are general-purpose and not customized to work on specific, proprietary technology. An alternative would be to leverage open models by training and using them locally. However, the large number of trainable parameters implies a conspicuous cost and environmental impact during the training phase. Also, such models impose hardware requirements for the inference phase (\ie when the LLM is used to generate recommendations). For example, DeepSeek Coder-33B would require, at inference time, $\sim$64GB of memory (VRAM), which is more than what most of modern laptops are equipped with. Certainly, this solution could be circumvented by deploying models for inference on local servers and using them as services, yet this implies costs, increased complexity, and network overhead.

One way of reducing models' requirements at inference time is \emph{quantization}.  In its general meaning, the term quantization refers to the process of approximating a large set of numbers with a smaller set. This happens, for example, in signal processing, when an audio (analog) signal is converted into its digital form using a given number of bits. The more bits, the better the sound quality, but the larger the resulting file.
In the context of LLM, quantization implies representing the neural network weights, usually expressed as 16-bit floating-point (\texttt{fp16}), with simpler values, \,  e.g., 8-bit integers --- \texttt{int8}.
The quantization procedure is performed on the already trained model which, in our context, could be one of the code-specific LLMs described above (\eg DeepSeek Coder). To do so, an additional (but cheap) calibration process is performed to best approximate the original model weights. Then, in the \emph{compression} phase, the original weights of the model are replaced by the learned approximations. This drastically reduces the memory footprint.

Wei \etal \cite{wei2023towards} have been the first investigating code generation via quantization, showing that by quantizing the weights of a 16B-parameters LLM (\ie CodeGen \cite{codegen}) to \texttt{int8} it was possible to reduce its memory footprint by 29\% with a negligible reduction in the code generation capabilities. Due to the fast pace at which LLM capabilities and quantization techniques are evolving, in this work we present a differentiated replication of the study conducted by Wei \etal \cite{wei2023towards}. We experiment with the latest advancements in model quantization techniques, which allows to push the compression to extreme quantization levels such as 2 bits per parameter. Also, we consider different types of calibration datasets to guide the quantization process (\ie the training data used to approximate the original weights), including code-specific ones. Last but not least, we focus on newer and larger LLM-based code generators, up to 34B parameters, and extend the study to two programming languages (Python and Java) instead of Python-only \cite{wei2023towards}. 

We consider two families of code-specific LLMs, namely CodeLlama \cite{roziere2023code} and DeepSeek Coder \cite{deepseekcoder}, experimenting with three of their variants: CodeLlama 7B, 13B, and 33B, and DeepSeek Coder 1.3B, 7B, and 33B. We apply on them the recently proposed Additive Quantization with Learned Multi-Codebooks (AQLM) \cite{egiazarian2024extreme} quantization techniques, which achieved new state-of-the-art results for Natural Language Processing (NLP) benchmarks at extreme quantization levels (\eg 2 bits). In particular, we experiment with quantizations at 8, 4, 3, and 2 bits. We assess the code generation capabilities of the baseline (\ie the original model) and of the quantized models on two benchmarks, MultiPL-E \cite{cassano:tse2023} and McEval \cite{chai2024mceval}. We focus on the Java and Python instances from the two benchmarks for a total of 414 implementation problems. We report the percentage of cases in which the subject models managed to correctly implement the required code at the first attempt. On top of that, we experiment with three different calibration datasets: (i) a general-purpose dataset with randomly sampled content (random dataset); (ii) a mixed dataset comprising 50\% GitHub code files and 50\% samples from Stack Overflow discussions; and (iii) a code-specific dataset containing only Java and Python code.

The obtained results show that, thanks to novel quantization techniques, such as AQLM, it is possible to quantize LLMs for the task of code generation down to 4 bits without any significant decrease in performance. Such a process results in a memory footprint reduced by 70\%, allowing the deployment of \textbf{very} large models also on hardware-constrained devices (and, in general, lowering the deployment cost). To the best of our knowledge, we are the first to demonstrate the feasibility of reducing precision below 4 bits, going as low as 3 and 2 bits, across multiple large code models and benchmarks. Indeed, for the first time, we show that using ad-hoc calibration datasets, especially those containing code samples and technical language, helps mitigating the performance degradation typically associated with such extreme quantization levels.
Finally, we found that 
larger code models are more resilient to information loss during extreme quantization.

Our study, on the one side, confirms the findings by Wei \etal \cite{wei2023towards} with larger models, multiple programming languages, and different quantization techniques. On the other side, and more importantly, it shows that the recent progress makes quantization even more attractive, due to the potential for extreme quantization. By pushing the precision down to 3 and 2 bits, we are laying the groundwork for more efficient code-specialized LLMs and the comprehension of the limits of quantization techniques in this context. All the code and data used in our study are publicly available \cite{replication}.

% !TEX root = ../main.tex
\section{Background and Related Work}
\label{sec:related}

\subsubsection*{Green AI for Software Engineering}
\label{sec:green-ai}
Casta{\~n}o \etal \cite{castano2023exploring} documented the correlation between the DL models' carbon footprint and various factors like model size and parameter count. The significant energy consumption and carbon emissions associated with LLMs have also been widely documented \cite{patterson2021carbon,strubell2020energy}, emphasizing the need for techniques aimed at making these models energy-efficient. 

Several approaches have been proposed in this direction, such as Parameter-Efficient Fine-Tuning (PEFT) \cite{houlsby2019parameter, hu2021lora}, Knowledge Distillation (KD) \cite{hsieh2023distilling, su2024distilled}, and Quantization \cite{gholami2022survey}. 

PEFT and KD typically require an additional training phase, which remains very expensive for large models.

\textbf{Quantization} reduces the memory footprint of large language models (LLMs) by representing model parameters in lower-precision formats, such as 8-bit integers, instead of standard 16- or 32-bit floating points \cite{gholami2022survey}. Two main strategies exist: Quantization-Aware Training (QAT) and Post-Training Quantization (PTQ). While \textbf{QAT} integrates quantization during model training \cite{esser2019learned} and has been adapted for smaller LLMs \cite{liu2023llm}, its high computational cost makes it impractical for large-scale models. \textbf{PTQ}, in contrast, converts pre-trained models to lower-precision formats with minimal training, typically using lightweight calibration datasets \cite{cai2020zeroq}. Due to these advantages, we focus only on PTQ techniques in this work.

\textbf{PTQ} approaches include weights-only quantization \cite{frantar2022gptq, shang2023pb, dettmers2023spqr}, and weight-activation quantization \cite{xiao2023smoothquant, dettmers2022gpt3}. For more aggressive compression (\( \leq 4 \)) bit, recent methods like  \textit{QuIP\#} \cite{tseng2024quip}, \textit{VPTQ} \cite{liu2024vptq} and \textit{AQLM} \cite{egiazarian2024extreme} have advanced the state of the art. AQLM, in particular, achieves Pareto-optimal performance in sub-3-bit scenarios through Multi-Codebook Quantization (MCQ), significantly outperforming prior techniques on natural language-related benchmarks.

In the software engineering domain, little research has been conducted in this direction. 
Recent studies have explored model compression techniques to enhance the efficiency of large language models (LLMs) for code-related tasks. \citet{sun2024neural} proposed a dynamic inference method (Stop\&Exit Controller) that accelerates code completion by skipping layers with minimal performance degradation. QLoRA, which combines 4-bit quantization with LoRA fine-tuning, has shown promise for code generation and summarization tasks \cite{weyssow2023exploring, afrin2025resource}. A broader evaluation of compression techniques—including pruning, distillation, and quantization—by \citet{d2024compression} confirmed quantization as the most effective for maintaining performance while reducing memory usage.

The most relevant prior work is by \citet{wei2023towards}, who investigated 8-bit quantization of large code models such as PLBART, CodeT5, and CodeGen, demonstrating improved energy efficiency with only marginal performance loss. Building on this, our study targets state-of-the-art models—CodeLlama and DeepSeek-Coder (up to 34B parameters)—and applies AQLM for extreme quantization down to 2 bits. Furthermore, motivated by findings from \citet{williams2024impact} on the influence of calibration data, we systematically evaluate different calibration datasets to assess their role in preserving model quality under aggressive quantization.

\subsubsection*{Additive Quantization of Language Models}
\label{sec:bac-aqlm}
Additive Quantization of Language Models (AQLM) employs Multi-Codebook Quantization (MCQ): A generalization of vector quantization (VQ) used in the approximate nearest neighbor search algorithms quantizes multiple vector dimensions simultaneously by learning ``codebooks''—sets of learnable candidate vectors for data encoding. In VQ, a database vector is divided into sub-groups, with each group encoded using a vector from a learned codebook. This process facilitates efficient computation of distances or dot products for similarity searches by exploiting the linearity of dot products.
Additive Quantization (AQ) \cite{babenko2014additive}, a popular MCQ algorithm, performs vector optimization in a way that for a linear layer with \(d_{\text{in}}\) input and \(d_{\text{out}}\) output features, characterized by its weight matrix \(W \in \mathbb{R}^{d_{\text{out}} \times d_{\text{in}}}\) and a set of calibration inputs \(X \in \mathbb{R}^{d_{\text{in}} \times n}\), the objective is to find a configuration of quantized weights \(\hat{W}\) that minimizes the squared error between the outputs of the original and compressed layers. AQ divides weight rows into groups of \(g\) consecutive elements and represents each group by summing vectors selected from multiple codebooks \(C_1, \ldots, C_M\), where each codebook contains \(2^B\) vectors for \(B\)-bit encoding. Each group's weight is quantized by combining a chosen vector from each codebook, encoded as a one-hot vector \(b_m\). 
The full-weight representation is then achieved by concatenating these quantized segments according to the following structure:
\[
W_{\text{ci}} = \bigoplus_{m=1}^M C_{m,b_{i,1,m}} \oplus \cdots \oplus \bigoplus_{m=1}^M C_{m,b_{i,g,d_{\text{in}},m}}
\]
where \(\oplus\) denotes concatenation and each \(b_{i,j,m}\) is a one-hot code that specifies the selection from the \(m\)-th codebook for the \(i\)-th output unit and the \(j\)-th group of input dimensions. 
\section{Design} \label{sec:design}
The \emph{goal} of this study is to conduct a differentiated replication of the work by Wei \etal \cite{wei2023towards} to investigate the impact of AQLM on the performance of LLMs when applied to the code generation task. Also, we aim at studying the impact of different types of calibration datasets on the quantization process, an aspect not covered in the original study \cite{wei2023towards}.
As previously explained, AQLM allows the exploration of extreme quantization levels, down to 2 bits, which may unlock new possibilities in the deployment of LLM-based code generators on constrained hardware environments.
The \emph{context} of the study consists of two state-of-the-art code LLMs, \ie CodeLlama \cite{roziere2023code} and DeepSeek-Coder \cite{deepseekcoder}, and two multilingual code generation benchmarks, \ie MultiPL-E \cite{cassano:tse2023} and McEval \cite{chai2024mceval}.

We aim at addressing the following research questions (RQs):

\textbf{RQ$_1$:} \emph{How does low-bit quantization affect the model's code generation ability?} In RQ$_1$, we aim to investigate the effect of low-bit quantization on the model's ability to generate code starting from a textual description. As compared to the work by \cite{wei2023towards}, which mostly focuses on 8-bit quantization, we experiment with 8, 4, 3, and 2-bit quantization. The goal is to carefully study the trade-off between loss in code generation accuracy and savings in terms of memory footprint. The RQ$_1$'s findings will provide insights into the extent to which LLM-based code generators can be compressed without significantly compromising their performance. \\
	
\textbf{RQ$_2$:} \emph{What impact does the calibration dataset have on model performance?} In RQ$_2$, we analyze the impact of the calibration dataset on model performance. The calibration dataset is used by the quantization technique to learn how to best approximate the original model's weights. We experiment with various types of calibration datasets, including code-specific ones, which may be more suitable in the context of code generation and help in minimizing the loss in performance. Such an analysis is particularly important for the quantization of models targeting software engineering tasks and has not been conducted in the work of Wei \etal \cite{wei2023towards} where the authors focus solely on examining the extent to which the size of the calibration dataset, built on CodeSearchNet \cite{husain:arxiv2019}, affects quantization performance.\\

\textbf{RQ$_3$:} \emph{How does extreme quantization affect model accuracy across different model sizes?}  
	In RQ$_1$ and RQ$_2$ we fix the size of the experimented models to 7B for both CodeLlama and DeepSeek Coder while experimenting with four quantization levels (8, 4, 3, and 2 bits). 
	In this research question, we are interested in determining the extent to which the findings of the previous RQs remain valid when quantizing LLMs featuring tens of billions of parameters. More specifically, in RQ$_3$ we consider models having different sizes (\ie 1B, 7B, 13B, 33B, and 34B) while fixing the quantization level to 2 bits.  

\subsection{Code-Specific LLMs Used in the Study}
We focus on two state-of-the-art code LLMs, namely CodeLlama and DeepSeek-Coder. Both these LLMs have shown their effectiveness when applied to code generation tasks \cite{li2023structured,coignion2024performance,ren2024reflectioncoder}.

\subsubsection{CodeLlama \cite{roziere2023code}}
CodeLlama is a family of open-source LLMs tailored for coding applications, built on the general-purpose Llama 2 \cite{touvron2023llama} LLM. CodeLlama features Llama 2 models further trained on a corpus of 500 billion tokens, including code and natural language. Multiple versions of these models are publicly accessible and are available in various sizes, with parameters ranging from 7B to 70B. Additionally, CodeLlama is available in three specialized formats: A general-purpose version for coding, an Instruct version optimized for instruction tuning, and a Python-specialized variant. CodeLlama has been selected as a representative model of code-related LLMs given its former successful applications in the automation of various code-related tasks  \cite{weyssow2023exploring, huang2024template, sun2024source}. In our study, we exploit the general-purpose CodeLlama versions featuring 7B, 13B, and 34B parameters. While we attempted to experiment with the 70B version, the computational cost would have been too high for our infrastructure. However, given the results obtained with the 34B version, we believe that similar trends would be observed with the 70B version as well.

\subsubsection{DeepSeek-Coder \cite{deepseekcoder}}
DeepSeek-Coder features open-source LLMs ranging in size from 1B to 33B parameters. Each model is available in two operational modes: Instruct (optimized for instruction tuning) and Base. These models have undergone extensive pre-training on 2 trillion tokens (including code-related documents) and have shown superior performance compared to many state-of-the-art LLMs for automating software engineering tasks. Notably, DeepSeek-Coder has outperformed much larger models like GPT-3.5 \cite{brown2020language}. Additionally, the mid-sized version of DeepSeek-Coder, with 6.7B parameters, has proven competitive with the 33B version of CodeLlama.
In our study, we use the Base version of DeepSeek-Coder in its sizes 1B, 7B, 13B, and 33B parameters. 

\subsection{Quantization Technique}
\label{sub:technique}

As previously explained, we rely on the state-of-the-art method for quantization proposed by Egiazarian \etal \cite{egiazarian2024extreme}, namely AQLM. To clarify the specifics of our study, let us briefly review the fundamental workings of AQLM. Given a codebook consisting of $2^{B}$ candidate vectors and \textit{M} being the number of available codebooks, AQLM encodes groups of \textit{g} consecutive weights from each model layer by expressing them as the sum of vectors selected from \textit{M} distinct codebooks. 

A model can be quantized at different levels of precision, measured in bits per parameter, based on the values assigned to \textit{g} (number of weights to quantize together), \textit{M} (number of codebooks), and \textit{B} (number of bits per codebook). The choice of these values may have an impact on quantization time, inference latency, and on the final memory footprint and accuracy of the quantized model. In this work we focus on memory footprint and accuracy. Inference latency is a different optimization objective, which we leave as future work.
The specific configurations we use for each experimented quantization level are reported in \tabref{tab:quantization_values}. A few clarifications are needed about these configurations. First, the values for the \textit{g}, \textit{M}, and \textit{B} parameters are derived from the targeted quantization level. For example, setting \textit{g} = 8, \textit{M} = 4, \textit{B} = 15 results in $\sim$8 bits per model's parameters, while lowering \textit{M} = 2 halves the precision to 4 bits. Second, given a parameters' configuration, the obtained precision slightly varies among the models. For example, CodeLlama 7B quantized using \textit{g}=8, \textit{M}=1, and \textit{B}=15, results in an average of  2.02 bits per parameter, compared to 1.92 bits per parameter for the 33B version. Finally, for the smallest model (\ie DeepSeek-Coder 1B), which is only used in the context of RQ$_3$ with the extreme 2-bit quantization, we had to lower \textit{B} to 14 (rather than 15) to obtain a quantization level close to 2 bits (\ie 2.05 bits per parameter).  

\begin{table}[htpb]
	\caption{AQLM configuration used for quantization\vspace{-0.3cm}}
	\label{tab:quantization_values}
	\centering
	\relsize{-1}
	\begin{tabular}{c|r}
		\hline
		\multicolumn{1}{c|}{\textbf{Precision}} & \multicolumn{1}{c}{\textbf{Values}} \\ \hline
		8 bit                          & \textit{g} = 8, \textit{M} = 4, \textit{B} = 15       \\
		4 bit                          & \textit{g} = 8, \textit{M} = 2, \textit{B} = 15       \\
		3 bit                          & \textit{g} = 8, \textit{M} = 2, \textit{B} = 12       \\
		2 bit                          & \textit{g} = 8, \textit{M} = 1, \textit{B} = 15       \\ \hline
	\end{tabular}
	\vspace{-0.3cm}
\end{table}

Besides the specific AQLM parameters' configuration, we had to create the calibration dataset used by the technique to minimize the quantization error (see \secref{sec:bac-aqlm}).

This dataset is usually composed of textual documents either sourced from the training set of the model being quantized or mined from publicly available sources \cite{williams2024impact}. Since the authors of CodeLlama and DeepSeek-Coder have not made their training datasets publicly available, we extracted calibration samples from RedPajama \cite{redpajama}, one of the largest open-source datasets. RedPajama features 1.2 trillion tokens collected from diverse sources, including Common Crawl \cite{commoncrawl}, C4 \cite{raffel2019exploring}, GitHub \cite{github}, Arxiv, Wikipedia, and StackExchange. Most of these sources are very likely to be part of the tuning datasets used for the experimented models.

We create three different calibration datasets since they will be the focus of RQ$_2$. For each of them, we selected 1,024 samples from RedPajama, each having a sequence length of 4,096 tokens, mimicking what was done by Egiazarian \etal \cite{egiazarian2024extreme} when presenting AQLM. This means that from each of the 1,024 documents selected from RedPajama (\eg a Wikipedia article), we only extract 4,096 consecutive tokens. The three variations of the calibration dataset differ in the ``nature'' of the documents they feature:

\textbf{Random dataset:} We randomly sampled the 1,024 samples from all the documents in the RedPajama dataset. Such a dataset simulates the choice of a ``general purpose'' calibration dataset, which may or may not be optimal in the context of code generation. \tabref{tab:random_datasets} shows the distribution of the selected samples for both families of models. Note that we had to build different calibration datasets for the two families of LLMs since they exploit different tokenizers. Thus, a document may feature $n$ tokens for CodeLlama and $m$ for DeepSeek-Coder, with $n \neq m$.
	
\textbf{Mixed dataset:} This dataset features 50\% of samples coming from GitHub code file instances, with an equal distribution of Python and Java code (since those are the two languages we experiment with), and the remaining 50\% of samples coming from Stack Overflow discussions having a question score greater than 10 and at least one reference to the terms ``Python" or ``Java". In both cases, the samples are again extracted from the RedPajama dataset, \eg we randomly select 512 GitHub files (256 Java and 256 Python) from RedPajama.
	
\textbf{Code dataset:} The whole 1,024 instances come from the GitHub subset of RedPajama, with 512 being Java files and 512 Python files.

\begin{table}[h]
    \centering
    \caption{Distribution of the Random dataset for the two LLMs\vspace{-0.3cm}}
    \label{tab:random_datasets}
	\relsize{-1}
    \begin{tabular}{l|r|r}
        \hline
        \textbf{Source} & \textbf{CodeLlama} & \textbf{DeepSeek-Coder} \\ \hline
        Common Crawl    & 834                & 835                    \\
        C4              & 76                 & 70                     \\
        Github          & 62                 & 61                     \\
        Arxiv           & 28                 & 30                     \\
        Wikipedia       & 20                 & 24                     \\
        StackExchange   & 4                  & 4                      \\ \hline
    \end{tabular}
    \vspace{-0.2cm}
\end{table}

\noindent We use the Random dataset to address RQ$_1$, while all three datasets are used to answer RQ$_2$. In RQ$_3$, we will exploit the dataset providing the best results as output of RQ$_2$.

\subsection{Evaluation Dataset}
Similarly to previous studies \cite{humaneval,wei2023towards,cassano2023knowledge}, we evaluate the model's code generation ability by providing a specification as input and expecting a valid implementation as output. More specifically, the code generation task considers a function's signature and documentation (\ie natural language description) as a specification and its body as the expected implementation.

We rely on two multilingual benchmarks: MultiPL-E \cite{cassano:tse2023} and McEval \cite{chai2024mceval}. The first includes two popular code generation datasets, \ie HumanEval \cite{humaneval} and MBPP \cite{mbpp}. Among these two datasets, we only used HumanEval, which was found to be more challenging than MBPP \cite{cassano:tse2023}. HumanEval originally included 164 hand-written Python prompts along with canonical solutions and unit test cases for evaluating model generations. The MultiPL-E benchmark retains 161 of these Python tasks, 158 of which have also been translated into Java. Note that three tasks were not translated because they used a Pythonic syntax not directly mappable into Java constructs. McEval is a collection of human-annotated coding tasks spanning different programming languages. Each problem has a signature, a docstring with an exhaustive description of the program requirements, a list of test cases, and a label indicating the level of difficulty (easy, middle, or hard) of the code to implement. We extracted the Python and Java problems from this benchmark, which amounted to 50 and 53 samples, respectively. We removed 8 problems from the Python samples that overlap with the HumanEval dataset, thus reducing the total number of Python samples to 42. The final Python (Java) problems feature 23 (30) easy tasks, 10 (13) middle-level samples, and 9 (10) hard problems. Our decision of focusing on Python and Java is due to their popularity and to the fact that they are considered high-resource languages, namely languages for which LLMs can exploit massive training material, thus usually exhibiting good code generation performances. This cannot really be said for low-resource languages (\eg R, Racket), which would have introduced an additional challenge for the LLMs besides the extreme quantization we experiment with.

\subsection{Study Procedure and Data Analysis}
\label{sub:analysis}
To answer RQ$_1$ we start by quantizing CodeLlama 7B and DeepSeek-Coder 7B via AQLM \cite{egiazarian2024extreme} by leveraging the ``Random'' calibration dataset. The quantization is performed at 8, 4, 3, and 2 bits following the configurations in \tabref{tab:quantization_values}. Note that we experiment with only the 7B versions to contain the computational cost of the study while still being able to observe the impact of quantization on reasonably sized models.
The resulting models and the baseline (\ie the non-quantized model) are then tested against the Java and Python samples from the two benchmarks to see how quantization affects model performance. In particular, given a code generation task, the candidate code (\ie the one generated by an LLM) is validated against task-specific unit tests and labeled as ``pass" if it satisfies all assertions or ``fail" otherwise.
As done by Wei \etal \cite{wei2023towards}, as a performance metric, we use \textit{pass@1}, which assesses the ability of the model to predict a correct solution with a single attempt \cite{humaneval} (\ie the model is only allowed to output a single candidate solution). To account for the nondeterminism of the LLM, we invoke the model 20 times per code generation problem using a temperature of 0.2 and a maximum output length of 1,024 tokens for MultiPL-E problems and 1,500 tokens for the McEval benchmark, ensuring an output window capable of handling the more demanding requirements of McEval.
As anticipated, at each of the 20 invocations, the model is only allowed to generate a single candidate solution. Finally, we compute the \textit{pass@1} score for each code generation task, \ie the percentage of cases (out of 20) in which the LLM produced a correct solution with the given single attempt. Finally, we report the average score across all code generation tasks. 

We compare the base model's \textit{pass@1} score on the Java and Python benchmarks to that of its quantized version. We complement this metric with statistical tests that consider the distribution of correct and incorrect predictions for each code generation task. In particular, we employ McNemar's test~\cite{mcnemar} for pairwise comparisons of dichotomous results from two distinct treatments and the Odds Ratio (OR) effect size to determine the magnitude of these differences. To adjust \emph{p}-values for multiple comparisons, we use the Benjamini-Hochberg procedure~\cite{yoav:jstor1995}.

To answer RQ$_2$, we quantized CodeLlama 7B and DeepSeek-Coder 7B again at the same quantization levels used in RQ$_1$ (\ie 8, 4, 3, and 2 bits) using, however, the two additional calibration datasets presented in \secref{sub:technique} (\ie ``Mixed'' and ``Code'' dataset). Given the goal of RQ$_2$ (\ie investigating the impact on the performance of different calibration datasets), the baseline is represented by the quantized models used in RQ$_1$, which exploited the ``Random'' calibration dataset. 
Thus, we compare the \textit{pass@1} achieved by the models quantized with the ``Mixed'' and ``Code'' calibration datasets against one of the models quantized with the ``Random'' dataset. We use the same statistical tests described for RQ$_1$.

As per RQ$_3$, we quantize models of different sizes (\eg  7B, 13B, and 34B versions of CodeLlama and 1B, 7B and 33B versions of DeepSeek-Coder) to the lowest precision, \ie 2 bits per parameter. We then observe how the performance drop between the base (non-quantized) model and the 2-bit quantized version varies with respect to the number of model's parameters, because we want to check the extent to which larger models are impacted by extreme quantization.
In RQ$_3$ we leverage the best-performing calibration dataset resulting from RQ$_2$ during the quantization procedure. Again, \textit{pass@1} score is our dependent variable, and we mirror the previously described statistical analysis also in this RQ.

Finally, in RQ$_1$ and RQ$_3$, we also experiment with an additional recently proposed strategy to further boost the performance of quantized models \cite{tseng2024quip,egiazarian2024extreme,leconte2024real}. Such a strategy, known as ``end-to-end fine-tuning'', can potentially extend the boundaries of the Pareto optimality, particularly when quantizing at low bit-size precision, such as 2 bits. The basic idea behind this additional fine-tuning (details in Appendix A of \cite{egiazarian2024extreme}) is to use the non-quantized LLM as a teacher model from which its quantized version can distill knowledge in order to minimize differences in their predictions on a given dataset. We run the end-to-end fine-tuning on the calibration dataset employed for the quantization process and re-used the same hyperparameter configuration by \cite{egiazarian2024extreme}: 5 training epochs, a learning rate of 1e-5 combined with the Adam optimizer \cite{zhang2018improved}, and an early stopping mechanism to prevent overfitting. We discuss the effect that this further ``optimization strategy'' has on the performance of quantized models at 3 and 2-bit levels. We do not present this analysis in RQ$_2$ since our only goal there is to isolate the impact of the type of calibration dataset. 

% !TEX root = ../main.tex
\section{Study Results}
\label{sec:results}
In this section, we report and discuss the results of our study to address the research questions formulated in \secref{sec:design}. When reporting statistical tests, we adopt the following notation for the significance level: • indicates a $p$-value $x$ being ($0.05 \leq x < 0.1$); 

\noindent
* ($0.01 \leq x < 0.05$); ** ($0.001 \leq x < 0.01$); and *** ($x < 0.001$).

\subsubsection*{How does low-bit quantization affect the model's code generation ability?}
\label{subsec:rq1}

\begin{table*}[tb]
    \caption{\textit{Pass@1} score of the quantized models in comparison to their corresponding baseline models\vspace{-0.4cm}}
     
    \centering
    \small
    \resizebox{0.76\linewidth}{!}{%
  \begin{tabular}{l|lr|lr|rrr|rrr}
    \toprule
    & Model & Params & Precision & Size & \multicolumn{3}{c|}{\textbf{Python}} & \multicolumn{3}{c}{\textbf{Java}} \\
    \cmidrule{6-11}
    & & & & & pass@1 & p-value & OR & pass@1 & p-value & OR \\
  \midrule

  \multirow{10}{*}{\rotatebox[origin=c]{90}{\textbf{MultiPL-E}}} & \multirow{2}*{\makecell[l]{CodeLlama - Base~\\\cite{roziere2023code}}}   & \multirow{2}*{7B}  & Float16 - Baseline & 13.48 GB & 29.8 & ------ & ------ & 32.2 & ------ & ------    \\
& &                                                      & 8-bit & 7.47 GB & 29.7 &  & 1.02 & 31.6 &  & 1.12    \\
& &                                                      & 4-bit & 4.00 GB & 29.1 &  & 1.17 & 30.7 & * & 1.29    \\
& &                                                      & 3-bit & 3.80 GB & 24.3 & *** & 2.88 & 26.5 & *** & 2.52    \\
& &                                                      & 2-bit & 2.26 GB & 16.4 & *** & 7.86 & 14.1 & *** & 20.83    \\\cline{2-11}
& \multirow{2}*{\makecell[l]{DeepSeek-Coder - Base\\\cite{deepseekcoder}}}   & \multirow{2}*{7B}  & Float16 - Baseline & 13.48 GB & 45.8 & ------ & ------ & 41.4 & ------ & ------    \\
& &                                                      & 8-bit & 7.48 GB & 46.2 &  & 0.93 & 41.9 &  & 0.92    \\
& &                                                      & 4-bit & 4.00 GB & 45.2 &  & 1.1 & 41.4 &  & 1.01    \\
& &                                                      & 3-bit & 3.80 GB & 41.1 & *** & 1.9 & 37.7 & *** & 1.69    \\
& &                                                      & 2-bit & 2.27 GB & 27.6 & *** & 7.89 & 23.2 & *** & 7.78    \\
\bottomrule
\multirow{10}{*}{\rotatebox[origin=c]{90}{\textbf{McEval}}} & \multirow{2}*{\makecell[l]{CodeLlama - Base~\\\cite{roziere2023code}}}   & \multirow{2}*{7B}  & Float16 - Baseline & 13.48 GB & 12.9 & ----- & ----- & 29.3 & ----- & -----    \\
& &                                                      & 8-bit & 7.47 GB & 12.9 &  & 1.0 & 29.2 &  & 1.02    \\
& &                                                      & 4-bit & 4.00 GB & 15.2 & • & 0.66 & 25.3 & *** & 1.96    \\
& &                                                      & 3-bit & 3.80 GB & 10.0 & * & 1.92 & 21.3 & *** & 2.85    \\
& &                                                      & 2-bit & 2.26 GB & 5.6 & *** & 3.44 & 11.4 & *** & 6.94    \\\cline{2-11}
& \multirow{2}*{\makecell[l]{DeepSeek-Coder - Base\\\cite{deepseekcoder}}}   & \multirow{2}*{7B}  & Float16 - Baseline & 13.48 GB & 41.8 & ----- & ----- & 42.6 & ----- & -----    \\
& &                                                      & 8-bit & 7.48 GB & 42.5 &  & 0.89 & 42.8 &  & 0.98    \\
& &                                                      & 4-bit & 4.00 GB & 40.7 &  & 1.2 & 45.9 & * & 0.7    \\
& &                                                      & 3-bit & 3.80 GB & 36.2 & *** & 2.21 & 34.5 & *** & 2.02    \\
& &                                                      & 2-bit & 2.27 GB & 13.7 & *** & 16.73 & 23.6 & *** & 3.22    \\
\bottomrule
\end{tabular}
\label{tab:result_rq1}
    }
    \vspace{-0.2cm}
\end{table*}

\tabref{tab:result_rq1} reports for each LLM subject of RQ$_1$ (\ie CodeLlama and DeepSeek-Coder 7B) and for both the baseline \texttt{fp16} precision model as well as for all its quantized versions: (i) their memory footprint in terms of GB, (ii) the \textit{pass@1} score they achieved on both the Python and the Java benchmarks; and (iii) the results of the statistical tests (\ie adjusted $p$-value and OR), in which we compare each quantized model against the baseline. For example, looking at CodeLlama 7B, the \texttt{fp16}-precision baseline requires 13.48GB of memory to achieve a 29.8\% of \textit{pass@1} score on the Python MultiPL-E benchmark, while its 4-bit quantized version only uses 4.00GB of memory with a \textit{pass@1} of 29.1\%. This difference in performance is not statistically significant. The 4-bit quantized model, when compared to the \texttt{fp16}-precision baseline, allows achieving a 70\% of memory reduction (\ie $\frac{13.48-4.00}{13.48}$) at the rather small cost of a relative -2\% (\ie $\frac{29.1-29.8}{29.8}$) decrease in \textit{pass@1} score.

Looking at \tabref{tab:result_rq1}, the first conclusion that can be made is that, independently from the model and programming language, 4-bit quantization is a safe bet, ensuring high memory reduction with very similar performance in terms of code generation. 
The result of the 4-bit CodeLlama model on the Java McEval benchmark is the only exception to this finding, with a statistically significant decrease of \textit{pass@1} by 4\%. To determine whether this behavior is due to the higher complexity of the benchmark, we checked the difficulty level of each prompt for which the base model returned a correct implementation while the quantized model did not. We discovered that 34 failed generations refer to \textit{easy} problems, 16 to \textit{middle-level} tasks, and 14 to \textit{hard} tasks, which is consistent with the distribution of problem levels in the benchmark (30 \textit{easy}, 10 \textit{middle}, and 12 \textit{hard} prompts). After a manual inspection, we found that the quantized model mainly failed due to assertion errors or calls to non-existent APIs.

When moving towards more extreme quantizations (\ie 3 and 2-bit), the price to pay becomes higher, with a statistically significant loss in \textit{pass@1} (see \tabref{tab:result_rq1}) which can reach a relative -67.22\% (2-bit quantization, DeepSeek Coder, McEval Python benchmark). Except for the previous case, DeepSeek-Coder achieves better overall performance than CodeLlama, both in terms of code generation capabilities at \texttt{fp16}-precision as well as in loss of performance with quantization, which is in most of cases lower on DeepSeek-Coder.

\begin{tcolorbox}[top=0pt,bottom=0pt,left=1pt,right=1pt,title=\faLightbulbO~Findings]
	\small
	With a 4-bit quantization, it is possible to reduce memory footprint by 70\% while preserving the code generation capabilities of the LLM. At more extreme quantization levels, instead, the models suffer major decreases in performance. The work we partially replicate \cite{wei2023towards} concluded that 8-bit was a safe choice. Our findings, achieved with a more recent quantization technique, push the boundaries further down.
	\normalsize
\end{tcolorbox}

As anticipated in \secref{sub:analysis} we also experimented with end-to-end fine-tuning to boost the performances at 3 and 2-bit quantization, as suggested by Egiazarian \etal \cite{egiazarian2024extreme}. \tabref{tab:results_rq1_finetuned} presents a comparison of the \textit{pass@1} score achieved by the 3- and 2-bit precision models, both with (dark grey) and without (light grey) end-to-end fine-tuning. For reference, the \texttt{fp16}-precision baseline is reported as well. In this case, the results of the statistical tests refer to the comparison between a quantized model with end-to-end fine-tuning versus the same quantized models without end-to-end fine-tuning. In other words, a statistically significant $p$-value indicates a significant boost of \textit{pass@1} score given by the additional fine-tuning. 

Our findings show that post-quantization fine-tuning can help in significantly boosting the performance of the 2-bit quantized models, with an average relative increase in the \textit{pass@1} score of 29.31\% for CodeLlama and 24.23\% for DeepSeek-Coder. In the case of 3-bit models, no significant differences are observed. Still, the 2-bit models, despite the significant increase in performance, exhibit considerable degradation when compared to their \texttt{fp16}-precision versions. 

\begin{table*}[htpb]
    
    \caption{Pass@1 accuracy of the quantized models compared to their fine-tuned versions\vspace{-0.4cm}}
    \centering
    \small
    \resizebox{0.8\linewidth}{!}{%
  \begin{tabular}{l|lr|lr|rrr|rrr}
    \toprule
    & Model & Params & Precision & Size & \multicolumn{3}{c|}{\textbf{Python}} & \multicolumn{3}{c}{\textbf{Java}} \\
    \cmidrule{6-11}
    & & & & & pass@1 & p-value & OR & pass@1 & p-value & OR \\
  \midrule

  \multirow{10}{*}{\rotatebox[origin=c]{90}{\textbf{MultiPL-E}}} & \multirow{2}*{\makecell[l]{CodeLlama - Base~\\\cite{roziere2023code}}}   & \multirow{2}*{7B}  & Float16 - Baseline & 13.48 GB & 29.8 & ------ & ------ & 32.2 & ------ & ------    \\
& &                                                      & \verylightgray{3-bit} & \verylightgray{3.80 GB} & \verylightgray{24.3} & \verylightgray{------} & \verylightgray{------} & \verylightgray{26.5} & \verylightgray{------} & \verylightgray{------}    \\
& &                                                      & \verylightgray{2-bit} & \verylightgray{2.26 GB} & \verylightgray{16.4} & \verylightgray{------} & \verylightgray{------} & \verylightgray{14.1} & \verylightgray{------} & \verylightgray{------}    \\
& &                                                      & \gray{3-bit + Fine-tuning} & \gray{3.80 GB} & \color{red}\raisebox{0ex}{$\blacktriangledown$} \gray{\textcolor{black}{\textbf{24.0}}} & \gray{} & \gray{0.91} & \color{green}\raisebox{0ex}{$\blacktriangle$} \gray{\textcolor{black}{\textbf{27.8}}} & \gray{*} & \gray{1.31}    \\
& &                                                      & \gray{2-bit + Fine-tuning} & \gray{2.26 GB} & \color{green}\raisebox{0ex}{$\blacktriangle$} \gray{\textcolor{black}{\textbf{19.9}}} & \gray{***} & \gray{2.01} & \color{green}\raisebox{0ex}{$\blacktriangle$} \gray{\textcolor{black}{\textbf{19.0}}} & \gray{***} & \gray{2.87}    \\\cline{2-11}
& \multirow{2}*{\makecell[l]{DeepSeek-Coder - Base\\\cite{deepseekcoder}}}  & \multirow{2}*{7B}  & Float16 - Baseline & 13.48 GB & 45.8 & ------ & ------ & 41.4 & ------ & ------    \\
& &                                                      & \verylightgray{3-bit} & \verylightgray{3.80 GB} & \verylightgray{41.1} & \verylightgray{------} & \verylightgray{------} & \verylightgray{37.7} & \verylightgray{------} & \verylightgray{------}    \\
& &                                                      & \verylightgray{2-bit} & \verylightgray{2.27 GB} & \verylightgray{27.6} & \verylightgray{------} & \verylightgray{------} & \verylightgray{23.2} & \verylightgray{------} & \verylightgray{------}    \\
& &                                                      & \gray{3-bit + Fine-tuning} & \gray{3.80 GB} & \color{green}\raisebox{0ex}{$\blacktriangle$} \gray{\textcolor{black}{\textbf{41.8}}} & \gray{} & \gray{1.13} & \color{red}\raisebox{0ex}{$\blacktriangledown$} \gray{\textcolor{black}{\textbf{37.7}}} & \gray{} & \gray{0.99}    \\
& &                                                      & \gray{2-bit + Fine-tuning} & \gray{2.27 GB} & \color{green}\raisebox{0ex}{$\blacktriangle$} \gray{\textcolor{black}{\textbf{33.0}}} & \gray{***} & \gray{2.67} & \color{green}\raisebox{0ex}{$\blacktriangle$} \gray{\textcolor{black}{\textbf{26.8}}} & \gray{***} & \gray{1.6}    \\
\bottomrule
\multirow{10}{*}{\rotatebox[origin=c]{90}{\textbf{McEval}}} & \multirow{2}*{\makecell[l]{CodeLlama - Base~\\\cite{roziere2023code}}}   & \multirow{2}*{7B}  & Float16 - Baseline & 13.48 GB & 12.9 & ----- & ----- & 29.3 & ----- & -----    \\
& &                                                      & \lightgray{3-bit} & \lightgray{3.80 GB} & \lightgray{10.0} & \lightgray{-----} & \lightgray{-----} & \lightgray{21.3} & \lightgray{-----} & \lightgray{-----}    \\
& &                                                      & \lightgray{2-bit} & \lightgray{2.26 GB} & \lightgray{5.6} & \lightgray{-----} & \lightgray{-----} & \lightgray{11.4} & \lightgray{-----} & \lightgray{-----}    \\
& &                                                      & \gray{3-bit + Fine-tuning} & \gray{3.80 GB} & \color{green}\raisebox{0ex}{$\blacktriangle$} \gray{\textcolor{black}{{\textbf{10.8}}}} & \gray{} & \gray{1.37} & \color{green}\raisebox{0ex}{$\blacktriangle$} \gray{\textcolor{black}{{\textbf{22.0}}}} & \gray{} & \gray{1.13}    \\
& &                                                      & \gray{2-bit + Fine-tuning} & \gray{2.26 GB} & \color{green}\raisebox{0ex}{$\blacktriangle$} \hspace{0.5em}\gray{\textcolor{black}{{\textbf{7.6}}}} & \gray{*} & \gray{1.81} & \color{green}\raisebox{0ex}{$\blacktriangle$} \gray{\textcolor{black}{{\textbf{14.3}}}} & \gray{**} & \gray{1.84}    \\\cline{2-11}
& \multirow{2}*{\makecell[l]{DeepSeek-Coder - Base\\\cite{deepseekcoder}}}  & \multirow{2}*{7B}  & Float16 - Baseline & 13.48 GB & 41.8 & ----- & ----- & 42.6 & ----- & -----    \\
& &                                                      & \lightgray{3-bit} & \lightgray{3.80 GB} & \lightgray{36.2} & \lightgray{-----} & \lightgray{-----} & \lightgray{34.5} & \lightgray{-----} & \lightgray{-----}    \\
& &                                                      & \lightgray{2-bit} & \lightgray{2.27 GB} & \lightgray{13.7} & \lightgray{-----} & \lightgray{-----} & \lightgray{23.6} & \lightgray{-----} & \lightgray{-----}    \\
& &                                                      & \gray{3-bit + Fine-tuning} & \gray{3.80 GB} & \color{red}\raisebox{0ex}{$\blacktriangledown$} \hspace{0.25em}\gray{\textcolor{black}{{\gray{35.6}}}} & \gray{} & \gray{0.9} & \color{red}\raisebox{0ex}{$\blacktriangledown$} \hspace{0.25em}\gray{\textcolor{black}{{\gray{32.4}}}} & \gray{} & \gray{0.79}    \\
& &                                                      & \gray{2-bit + Fine-tuning} & \gray{2.27 GB} & \color{green}\raisebox{0ex}{$\blacktriangle$} \gray{\textcolor{black}{{\textbf{20.2}}}} & \gray{***} & \gray{2.38} & \color{green}\raisebox{0ex}{$\blacktriangle$} \gray{\textcolor{black}{{\textbf{27.0}}}} & \gray{**} & \gray{1.49}    \\
\bottomrule
\end{tabular}
\label{tab:results_rq1_finetuned}
    }
    \vspace{-0.2cm}
\end{table*}

\begin{tcolorbox}[top=0pt,bottom=0pt,left=1pt,right=1pt,title=\faLightbulbO~Findings]
	\small
	Fine-tuning after quantization can help in boosting the performance of 2-bit quantized models, while it does not really help already at 3-bit precision. In summary, 4-bit is still the recommended quantization level given the current state-of-the-art.
	\normalsize
\end{tcolorbox}

\begin{table*}[htpb]
    
    \caption{Pass@1 accuracy of the quantized models using different calibration datasets\vspace{-0.4cm}}
    \centering
    \small
    \resizebox{0.88\linewidth}{!}{%
  \begin{tabular}{l|lr|lr|rrr|rrr}
    \toprule
    & Model & Params & Precision & Size & \multicolumn{3}{c|}{\textbf{Python}} & \multicolumn{3}{c}{\textbf{Java}} \\
    \cmidrule{6-11}
    & & & & & pass@1 & p-value & OR & pass@1 & p-value & OR \\
  \midrule

  \multirow{28}{*}{\rotatebox[origin=c]{90}{\textbf{MultiPL-E}}} & \multirow{2}*{\makecell[l]{CodeLlama - Base~\\\cite{roziere2023code}}}   & \multirow{2}*{7B}  & Float16 - Baseline & 13.48 GB & 29.8 & ------ & ------ & 32.2 & ------ & ------    \\
& &                                                      & \verylightgray{8-bit with Random samples} & \verylightgray{7.47 GB} & \verylightgray{29.7} & \verylightgray{------} & \verylightgray{------} & \verylightgray{31.6} & \verylightgray{------} & \verylightgray{------}    \\
& &                                                      & \verylightgray{8-bit with Mixed samples} & \verylightgray{7.47 GB} & \color{red}\raisebox{0ex}{$\blacktriangledown$} \verylightgray{\textcolor{black}{\textbf{29.7}}} & \verylightgray{} & \verylightgray{0.99} & \color{green}\raisebox{0ex}{$\blacktriangle$} \verylightgray{\textcolor{black}{\textbf{32.3}}} & \verylightgray{} & \verylightgray{1.13}    \\
& &                                                      & \verylightgray{8-bit with Code samples} & \verylightgray{7.47 GB} & \color{red}\raisebox{0ex}{$\blacktriangledown$} \verylightgray{\textcolor{black}{\textbf{29.2}}} & \verylightgray{} & \verylightgray{0.88} & \color{green}\raisebox{0ex}{$\blacktriangle$} \verylightgray{\textcolor{black}{\textbf{32.0}}} & \verylightgray{} & \verylightgray{1.08}    \\
& &                                                      & \lightgray{4-bit with Random samples} & \lightgray{4.00 GB} & \lightgray{29.1} & \lightgray{------} & \lightgray{------} & \lightgray{30.7} & \lightgray{------} & \lightgray{------}    \\
& &                                                      & \lightgray{4-bit with Mixed samples} & \lightgray{4.00 GB} & \color{red}\raisebox{0ex}{$\blacktriangledown$} \lightgray{\textcolor{black}{\textbf{29.0}}} & \lightgray{} & \lightgray{0.96} & \color{green}\raisebox{0ex}{$\blacktriangle$} \lightgray{\textcolor{black}{\textbf{31.4}}} & \lightgray{} & \lightgray{1.12}    \\
& &                                                      & \lightgray{4-bit with Code samples} & \lightgray{4.00 GB} & \color{green}\raisebox{0ex}{$\blacktriangle$} \lightgray{\textcolor{black}{\textbf{30.2}}} & \lightgray{} & \lightgray{1.22} & \color{red}\raisebox{0ex}{$\blacktriangledown$} \lightgray{\textcolor{black}{\textbf{29.8}}} & \lightgray{} & \lightgray{0.85}    \\
& &                                                      & \gray{3-bit with Random samples} & \gray{3.80 GB} & \gray{24.3} & \gray{------} & \gray{------} & \gray{26.5} & \gray{------} & \gray{------}    \\
& &                                                      & \gray{3-bit with Mixed samples} & \gray{3.80 GB} & \color{green}\raisebox{0ex}{$\blacktriangle$} \gray{\textcolor{black}{\textbf{28.2}}} & \gray{***} & \gray{2.16} & \color{green}\raisebox{0ex}{$\blacktriangle$} \gray{\textcolor{black}{\textbf{28.4}}} & \gray{**} & \gray{1.38}    \\
& &                                                      & \gray{3-bit with Code samples} & \gray{3.80 GB} & \color{green}\raisebox{0ex}{$\blacktriangle$} \gray{\textcolor{black}{\textbf{27.0}}} & \gray{***} & \gray{1.95} & \color{green}\raisebox{0ex}{$\blacktriangle$} \gray{\textcolor{black}{\textbf{28.0}}} & \gray{*} & \gray{1.28}    \\
& &                                                      & \darkgray{2-bit with Random samples} & \darkgray{2.26 GB} & \darkgray{16.4} & \darkgray{------} & \darkgray{------} & \darkgray{14.1} & \darkgray{------} & \darkgray{------}    \\
& &                                                      & \darkgray{2-bit with Mixed samples} & \darkgray{2.26 GB} & \color{green}\raisebox{0ex}{$\blacktriangle$} \darkgray{\textcolor{black}{\textbf{23.9}}} & \darkgray{***} & \darkgray{3.95} & \color{green}\raisebox{0ex}{$\blacktriangle$} \darkgray{\textcolor{black}{\textbf{21.5}}} & \darkgray{***} & \darkgray{4.21}    \\
& &                                                      & \darkgray{2-bit with Code samples} & \darkgray{2.26 GB} & \color{green}\raisebox{0ex}{$\blacktriangle$} \darkgray{\textcolor{black}{\textbf{24.1}}} & \darkgray{***} & \darkgray{4.7} & \color{green}\raisebox{0ex}{$\blacktriangle$} \darkgray{\textcolor{black}{\textbf{19.4}}} & \darkgray{***} & \darkgray{2.51}    \\\cline{2-11}
& \multirow{2}*{\makecell[l]{DeepSeek-Coder - Base\\\cite{deepseekcoder}}}   & \multirow{2}*{7B}  & Float16 - Baseline & 13.48 GB & 45.8 & ------ & ------ & 41.4 & ------ & ------    \\
& &                                                      & \verylightgray{8-bit with Random samples} & \verylightgray{7.48 GB} & \verylightgray{46.2} & \verylightgray{------} & \verylightgray{------} & \verylightgray{41.9} & \verylightgray{------} & \verylightgray{------}    \\
& &                                                      & \verylightgray{8-bit with Mixed samples} & \verylightgray{7.48 GB} & \color{red}\raisebox{0ex}{$\blacktriangledown$} \verylightgray{\textcolor{black}{\textbf{45.4}}} & \verylightgray{} & \verylightgray{0.87} & \color{green}\raisebox{0ex}{$\blacktriangle$} \verylightgray{\textcolor{black}{\textbf{43.2}}} & \verylightgray{•} & \verylightgray{1.25}    \\
& &                                                      & \verylightgray{8-bit with Code samples} & \verylightgray{7.48 GB} &  \color{red}\raisebox{0ex}{$\blacktriangledown$} \verylightgray{\textcolor{black}{\textbf{45.9}}} & \verylightgray{} & \verylightgray{0.94} & \color{red}\raisebox{0ex}{$\blacktriangledown$} \verylightgray{\textcolor{black}{\textbf{41.7}}} & \verylightgray{} & \verylightgray{0.96}    \\
& &                                                      & \lightgray{4-bit with Random samples} & \lightgray{4.00 GB} & \lightgray{45.2} & \lightgray{------} & \lightgray{------} & \lightgray{41.4} & \lightgray{------} & \lightgray{------}    \\
& &                                                      & \lightgray{4-bit with Mixed samples} & \lightgray{4.00 GB} &  \color{red}\raisebox{0ex}{$\blacktriangledown$} \lightgray{\textcolor{black}{\textbf{44.5}}} & \lightgray{} & \lightgray{0.89} & \color{green}\raisebox{0ex}{$\blacktriangle$} \lightgray{\textcolor{black}{\textbf{41.8}}} & \lightgray{} & \lightgray{1.07}    \\
& &                                                      & \lightgray{4-bit with Code samples} & \lightgray{4.00 GB} & \color{red}\raisebox{0ex}{$\blacktriangledown$} \lightgray{\textcolor{black}{\textbf{44.2}}} & \lightgray{} & \lightgray{0.86} &  \color{red}\raisebox{0ex}{$\blacktriangledown$} \lightgray{\textcolor{black}{\textbf{40.6}}} & \lightgray{} & \lightgray{0.89}    \\
& &                                                      & \gray{3-bit with Random samples} & \gray{3.80 GB} & \gray{41.1} & \gray{------} & \gray{------} & \gray{37.7} & \gray{------} & \gray{------}    \\
& &                                                      & \gray{3-bit with Mixed samples} & \gray{3.80 GB} & \color{green}\raisebox{0ex}{$\blacktriangle$} \gray{\textcolor{black}{\textbf{43.7}}} & \gray{***} & \gray{1.4} & \color{green}\raisebox{0ex}{$\blacktriangle$} \gray{\textcolor{black}{\textbf{39.1}}} & \gray{} & \gray{1.23}    \\
& &                                                      & \gray{3-bit with Code samples} & \gray{3.80 GB} & \color{green}\raisebox{0ex}{$\blacktriangle$} \gray{\textcolor{black}{\textbf{42.5}}} & \gray{•} & \gray{1.2} & \color{green}\raisebox{0ex}{$\blacktriangle$} \gray{\textcolor{black}{\textbf{38.7}}} & \gray{} & \gray{1.14}    \\
& &                                                      & \darkgray{2-bit with Random samples} & \darkgray{2.27 GB} & \darkgray{27.6} & \darkgray{------} & \darkgray{------} & \darkgray{23.2} & \darkgray{------} & \darkgray{------}    \\
& &                                                      & \darkgray{2-bit with Mixed samples} & \darkgray{2.27 GB} & \color{green}\raisebox{0ex}{$\blacktriangle$} \darkgray{\textcolor{black}{\textbf{35.7}}} & \darkgray{***} & \darkgray{3.05} & \color{green}\raisebox{0ex}{$\blacktriangle$} \darkgray{\textcolor{black}{\textbf{27.4}}} & \darkgray{***} & \darkgray{1.66}    \\
& &                                                      & \darkgray{2-bit with Code samples} & \darkgray{2.27 GB} & \color{green}\raisebox{0ex}{$\blacktriangle$} \darkgray{\textcolor{black}{\textbf{34.8}}} & \darkgray{***} & \darkgray{2.57} & \color{green}\raisebox{0ex}{$\blacktriangle$} \darkgray{\textcolor{black}{\textbf{27.5}}} & \darkgray{***} & \darkgray{1.8}    \\
\bottomrule
\multirow{28}{*}{\rotatebox[origin=c]{90}{\textbf{McEval}}} & \multirow{2}*{\makecell[l]{CodeLlama - Base~\\\cite{roziere2023code}}}   & \multirow{2}*{7B}  & Float16 - Baseline & 13.48 GB & 12.9 & ----- & ----- & 29.3 & ----- & -----    \\
& &                                                      & \verylightgray{8-bit with Random samples} & \verylightgray{7.47 GB} & \verylightgray{12.9} & \verylightgray{-----} & \verylightgray{-----} & \verylightgray{29.2} & \verylightgray{-----} & \verylightgray{-----}    \\
& &                                                      & \verylightgray{8-bit with Mixed samples} & \verylightgray{7.47 GB} & \color{green}\raisebox{0ex}{$\blacktriangle$} \gray{\textcolor{black}{\textbf{\verylightgray{13.7}}}} & \verylightgray{} & \verylightgray{1.19} & \color{red}\raisebox{0ex}{$\blacktriangledown$} \gray{\textcolor{black}{\textbf{\verylightgray{28.6}}}} & \verylightgray{} & \verylightgray{0.88}    \\
& &                                                      & \verylightgray{8-bit with Code samples} & \verylightgray{7.47 GB} & \color{red}\raisebox{0ex}{$\blacktriangledown$} \gray{\textcolor{black}{\textbf{\verylightgray{12.3}}}} & \verylightgray{} & \verylightgray{0.88} & \color{green}\raisebox{0ex}{$\blacktriangle$} \gray{\textcolor{black}{\textbf{\verylightgray{29.5}}}} & \verylightgray{} & \verylightgray{1.05}    \\
& &                                                      & \lightgray{4-bit with Random samples} & \lightgray{4.00 GB} & \lightgray{15.2} & \lightgray{-----} & \lightgray{-----} & \lightgray{25.3} & \lightgray{-----} & \lightgray{-----}    \\
& &                                                      & \lightgray{4-bit with Mixed samples} & \lightgray{4.00 GB} & \color{red}\raisebox{0ex}{$\blacktriangledown$} \gray{\textcolor{black}{\textbf{\lightgray{13.0}}}} & \lightgray{•} & \lightgray{0.62} & \color{green}\raisebox{0ex}{$\blacktriangle$} \gray{\textcolor{black}{\textbf{\lightgray{30.3}}}} & \lightgray{***} & \lightgray{2.2}    \\
& &                                                      & \lightgray{4-bit with Code samples} & \lightgray{4.00 GB} & \color{red}\raisebox{0ex}{$\blacktriangledown$} \gray{\textcolor{black}{\textbf{\lightgray{11.1}}}} & \lightgray{***} & \lightgray{0.43} & \color{green}\raisebox{0ex}{$\blacktriangle$} \gray{\textcolor{black}{\textbf{\lightgray{25.8}}}} & \lightgray{} & \lightgray{1.1}    \\
& &                                                      & \gray{3-bit with Random samples} & \gray{3.80 GB} & \gray{10.0} & \gray{-----} & \gray{-----} & \gray{21.3} & \gray{-----} & \gray{-----}    \\
& &                                                      & \gray{3-bit with Mixed samples} & \gray{3.80 GB} & \color{green}\raisebox{0ex}{$\blacktriangle$} \gray{\textcolor{black}{\textbf{\gray{12.3}}}} & \gray{•} & \gray{1.86} & \color{green}\raisebox{0ex}{$\blacktriangle$} \gray{\textcolor{black}{\textbf{\gray{25.5}}}} & \gray{***} & \gray{1.92}    \\
& &                                                      & \gray{3-bit with Code samples} & \gray{3.80 GB} & \color{green}\raisebox{0ex}{$\blacktriangle$} \gray{\textcolor{black}{\textbf{\gray{10.8}}}} & \gray{} & \gray{1.35} & \color{red}\raisebox{0ex}{$\blacktriangledown$} \gray{\textcolor{black}{\textbf{\gray{19.9}}}} & \gray{} & \gray{0.77}    \\
& &                                                      & \darkgray{2-bit with Random samples} & \darkgray{2.26 GB} & \darkgray{5.6} & \darkgray{-----} & \darkgray{-----} & \darkgray{11.4} & \darkgray{-----} & \darkgray{-----}    \\
& &                                                      & \darkgray{2-bit with Mixed samples} & \darkgray{2.26 GB} & \color{green}\raisebox{0ex}{$\blacktriangle$} \gray{\textcolor{black}{\textbf{\darkgray{11.1}}}} & \darkgray{***} & \darkgray{4.54} & \color{green}\raisebox{0ex}{$\blacktriangle$} \gray{\textcolor{black}{\textbf{\darkgray{12.8}}}} & \darkgray{} & \darkgray{1.37}    \\
& &                                                      & \darkgray{2-bit with Code samples} & \darkgray{2.26 GB} & \color{green}\raisebox{0ex}{$\blacktriangle$} \hspace{0.5em}\gray{\textcolor{black}{\textbf{\darkgray{6.1}}}} & \darkgray{} & \darkgray{1.17} & \color{green}\raisebox{0ex}{$\blacktriangle$} \gray{\textcolor{black}{\textbf{\darkgray{12.8}}}} & \darkgray{} & \darkgray{1.36}    \\\cline{2-11}
& \multirow{2}*{\makecell[l]{DeepSeek-Coder - Base\\\cite{deepseekcoder}}}   & \multirow{2}*{7B}  & Float16 - Baseline & 13.48 GB & 41.8 & ----- & ----- & 42.6 & ----- & -----    \\
& &                                                      & \verylightgray{8-bit with Random samples} & \verylightgray{7.48 GB} & \verylightgray{42.5} & \verylightgray{-----} & \verylightgray{-----} & \verylightgray{42.8} & \verylightgray{-----} & \verylightgray{-----}    \\
& &                                                      & \verylightgray{8-bit with Mixed samples} & \verylightgray{7.48 GB} & \color{green}\raisebox{0ex}{$\blacktriangle$} \gray{\textcolor{black}{\textbf{\verylightgray{42.7}}}} & \verylightgray{} & \verylightgray{1.04} & \color{red}\raisebox{0ex}{$\blacktriangledown$} \gray{\textcolor{black}{\textbf{\verylightgray{42.5}}}} & \verylightgray{} & \verylightgray{0.97}    \\
& &                                                      & \verylightgray{8-bit with Code samples} & \verylightgray{7.48 GB} & \color{red}\raisebox{0ex}{$\blacktriangledown$} \gray{\textcolor{black}{\textbf{\verylightgray{41.3}}}} & \verylightgray{} & \verylightgray{0.83} & \color{red}\raisebox{0ex}{$\blacktriangledown$} \gray{\textcolor{black}{\textbf{\verylightgray{42.7}}}} & \verylightgray{} & \verylightgray{0.99}    \\
& &                                                      & \lightgray{4-bit with Random samples} & \lightgray{4.00 GB} & \lightgray{40.7} & \lightgray{-----} & \lightgray{-----} & \lightgray{45.9} & \lightgray{-----} & \lightgray{-----}    \\
& &                                                      & \lightgray{4-bit with Mixed samples} & \lightgray{4.00 GB} & \color{red}\raisebox{0ex}{$\blacktriangledown$} \gray{\textcolor{black}{\textbf{\lightgray{39.0}}}} & \lightgray{} & \lightgray{0.79} & \color{red}\raisebox{0ex}{$\blacktriangledown$} \gray{\textcolor{black}{\textbf{\lightgray{42.8}}}} & \lightgray{*} & \lightgray{0.74}    \\
& &                                                      & \lightgray{4-bit with Code samples} & \lightgray{4.00 GB} & \color{red}\raisebox{0ex}{$\blacktriangledown$} \gray{\textcolor{black}{\textbf{\lightgray{39.8}}}} & \lightgray{} & \lightgray{0.86} & \color{green}\raisebox{0ex}{$\blacktriangle$} \gray{\textcolor{black}{\textbf{\lightgray{46.3}}}} & \lightgray{} & \lightgray{1.04}    \\
& &                                                      & \gray{3-bit with Random samples} & \gray{3.80 GB} & \gray{36.2} & \gray{-----} & \gray{-----} & \gray{34.5} & \gray{-----} & \gray{-----}    \\
& &                                                      & \gray{3-bit with Mixed samples} & \gray{3.80 GB} & \color{red}\raisebox{0ex}{$\blacktriangledown$} \gray{\textcolor{black}{\textbf{\gray{35.5}}}} & \gray{} & \gray{0.92} & \color{green}\raisebox{0ex}{$\blacktriangle$} \gray{\textcolor{black}{\textbf{\gray{42.8}}}} & \gray{***} & \gray{2.19}    \\
& &                                                      & \gray{3-bit with Code samples} & \gray{3.80 GB} & \color{green}\raisebox{0ex}{$\blacktriangle$} \gray{\textcolor{black}{\textbf{\gray{36.5}}}} & \gray{} & \gray{1.06} & \color{green}\raisebox{0ex}{$\blacktriangle$} \gray{\textcolor{black}{\textbf{\gray{45.6}}}} & \gray{***} & \gray{2.43}    \\
& &                                                      & \darkgray{2-bit with Random samples} & \darkgray{2.27 GB} & \darkgray{13.7} & \darkgray{-----} & \darkgray{-----} & \darkgray{23.6} & \darkgray{-----} & \darkgray{-----}    \\
& &                                                      & \darkgray{2-bit with Mixed samples} & \darkgray{2.27 GB} & \color{green}\raisebox{0ex}{$\blacktriangle$} \gray{\textcolor{black}{\textbf{\darkgray{26.2}}}} & \darkgray{***} & \darkgray{4.39} & \color{green}\raisebox{0ex}{$\blacktriangle$} \gray{\textcolor{black}{\textbf{\darkgray{29.1}}}} & \darkgray{**} & \darkgray{1.55}    \\
& &                                                      & \darkgray{2-bit with Code samples} & \darkgray{2.27 GB} & \color{green}\raisebox{0ex}{$\blacktriangle$} \gray{\textcolor{black}{\textbf{\darkgray{24.6}}}} & \darkgray{***} & \darkgray{4.54} & \color{green}\raisebox{0ex}{$\blacktriangle$} \gray{\textcolor{black}{\textbf{\darkgray{28.0}}}} & \darkgray{**} & \darkgray{1.55}    \\
\bottomrule
\end{tabular}
\label{tab:resultrq2}
    }
    \vspace{-0.2cm}
\end{table*}

\subsubsection*{Which impact does the calibration dataset have on model performance?}
In RQ$_2$, we analyze whether the calibration dataset used at quantization time plays a major role in the code generation capabilities of the quantized LLMs. Specifically, we compare the \textit{pass@1} scores achieved in the previous experiment, \ie using the ``Random'' calibration dataset, with those obtained with the ``Mixed'' and ``Code'' datasets. \tabref{tab:resultrq2} reports the results of the three treatments. 

In \tabref{tab:resultrq2} the statistical tests compare each model quantized with the ``Mixed'' or with the ``Code'' dataset to its same version quantized using the ``Random'' dataset (\eg CodeLlama 8-bit with Mixed dataset \emph{vs} CodeLlama 8-bit with Random dataset).

For 8-bit and 4-bit models, we do not observe a statistically significant difference in \textit{pass@1} score between models quantized with different calibration datasets. This holds both for CodeLlama and DeepSeek-Coder and in both languages (Python and Java). This finding reveals that code models are robust to the calibration dataset provided at quantization time for target precisions greater or equal to 4 bits per parameter. The only exceptions are the CodeLlama 4-bit model on the McEval Java benchmark and the DeepSeek-Coder 4-bit model on the McEval Python benchmark, which are compared with a ``Random'' quantization that already performs better than the \texttt{fp16}-precision baseline. The 4-bit CodeLlama model, instead, shows a statistically significant improvement on the McEval Python benchmark after the ``Mixed'' calibration, thus recovering from the performance gap with the baseline observed in \secref{subsec:rq1}. 
On the other hand, we notice that 3-bit and 2-bit precision models are more sensible to the samples provided in the calibration dataset. Indeed, both models show a statistically significant improvement in performance when calibration datasets feature code samples, suggesting their need to better approximate the model weights at extreme quantization levels. Thanks to code-related calibration datasets, a 3-bit quantization might be considered an option in very resource-constrained hardware devices. However, when comparing the 3-bit models quantized using the ``Mixed'' dataset against the \texttt{fp16}-precision baseline, the difference in \textit{pass@1} score is still statistically significant.

Similarly, the 2-bit models continue to suffer from a significant performance gap despite major improvements brought by the code-related calibration datasets. This calls for additional research on optimizing calibration datasets for code-related tasks, which could help enhance performance even further. Indeed, we only experimented with three variants of the calibration datasets, which, due to the unavailability of the training sets used for CodeLlama and DeepSeek-Coder, we cannot guarantee to be representative of the training instances (a condition that is expected to help the quantization process \cite{egiazarian2024extreme}).
Although our results do not show a clear winner between the two datasets, providing both code and technical language generally yields better performance for both models. This is particularly evident in the CodeLlama family of models, as they are more susceptible to sophisticated natural language requirements, such as those found in the McEval dataset. We believe including code and domain-specific natural language in the calibration dataset can help models that frequently struggle to follow complex instructions, like CodeLlama.

\begin{tcolorbox}[top=0pt,bottom=0pt,left=1pt,right=1pt,title=\faLightbulbO~Findings]
	\small
	Calibration datasets significantly impact the performance of the quantized models when targeting precision levels lower than 4 bits per parameter. Providing code-related samples improves the quantization process, resulting in more performant code generation models.
	\normalsize
\end{tcolorbox}

\subsubsection*{How does extreme quantization affect model accuracy across different model sizes?}
In RQ$_2$, we reported findings related to the quantization of CodeLlama and DeepSeek-Coder when fixing their size to 7B parameters. In RQ$_3$, we explore the impact of extreme quantization on models of different sizes, namely CodeLlama 7B, 13B, and 34B, and DeepSeek-Coder 1B, 7B, and 33B. We decided to focus only on a single quantization level (\ie 2-bit) since the cost of running these experiments with models up to 34B parameters is extremely high. Also, we already observed that a 4-bit quantization can preserve the LLM's code generation capabilities when using a 7B-parameter model. Thus, we wanted to investigate whether larger models could handle the most extreme quantization level. We adopt the ``Mixed'' calibration dataset in this RQ. \tabref{tab:results_rq3} reports the achieved results, with the statistical tests comparing the quantized models to the \texttt{fp16}-precision baseline and the ``Dec (\%)'' column showing the percentage relative decrease in \textit{pass@1} score.
Note that in this RQ, we use 2-bit quantization without and with end-to-end fine-tuning since we observed its benefits for the 2-bit precision in RQ$_1$.

\begin{table*}[htpb]
    
    \caption{Pass@1 accuracy of the quantized models on different sizes\vspace{-0.4cm}}
    \centering
    \small
    \resizebox{0.91\linewidth}{!}{%
  \begin{tabular}{l|lr|lr|rrrr|rrrr}
    \toprule
    & Model & Params & Precision & Size (GB) & \multicolumn{4}{c|}{\textbf{Python}} & \multicolumn{4}{c}{\textbf{Java}} \\
    \cmidrule{6-13}
    & & & & & pass@1 & p-value & OR & Dec (\%) & pass@1 & p-value & OR & Dec (\%)\\
  \midrule

  \multirow{18}{*}{\rotatebox[origin=c]{90}{\textbf{MultiPL-E}}} & \multirow{2}*{\makecell[l]{CodeLlama - Base~\\\cite{roziere2023code}}}   & \multirow{2}*{7B}  & Float16 - Baseline & 13.48 & 29.8 & ----- & ----- & ----- & 32.2 & ----- & ----- & -----    \\
& &                                                      & \verylightgray{2-bit} & \verylightgray{2.26} & \verylightgray{\textbf{23.9}} & \verylightgray{***} & \verylightgray{2.68} & \verylightgray{-19.8} & \verylightgray{\textbf{21.5}} & \verylightgray{***} & \verylightgray{4.38} & \verylightgray{-33.2}    \\
& &                                                      & \gray{2-bit + Finetuning} & \gray{2.26} & \gray{\textbf{25.5}} & \gray{***} & \gray{2.01} & \gray{-14.4} & \gray{\textbf{26.5}} & \gray{***} & \gray{1.98} & \gray{-17.7}    \\
\cline{4-13}
& \multirow{2}*{~}   & \multirow{2}*{13B}  & Float16 - Baseline & 24.25 & 34.3 & ----- & ----- & ----- & 38.3 & ----- & ----- & -----    \\
& &                                                      & \verylightgray{2-bit} & \verylightgray{3.98} & \verylightgray{\textbf{30.9}} & \verylightgray{***} & \verylightgray{1.47} & \verylightgray{-9.9} & \verylightgray{\textbf{27.7}} & \verylightgray{***} & \verylightgray{3.35} & \verylightgray{-27.7}    \\
& &                                                      & \gray{2-bit + Finetuning} & \gray{3.98} & \gray{\textbf{30.1}} & \gray{***} & \gray{1.66} & \gray{-12.2} & \gray{\textbf{32.8}} & \gray{***} & \gray{1.84} & \gray{-14.4}    \\
\cline{4-13}
& \multirow{2}*{~}   & \multirow{2}*{34B}  & Float16 - Baseline & 62.74 & 41.9 & ----- & ----- & ----- & 44.1 & ----- & ----- & -----    \\
& &                                                      & \verylightgray{2-bit} & \verylightgray{9.54} & \verylightgray{\textbf{37.1}} & \verylightgray{***} & \verylightgray{1.62} & \verylightgray{-11.5} & \verylightgray{\textbf{32.7}} & \verylightgray{***} & \verylightgray{3.64} & \verylightgray{-25.9}    \\
& &                                                      & \gray{2-bit + Finetuning} & \gray{9.54} & \gray{\textbf{36.0}} & \gray{***} & \gray{1.97} & \gray{-14.1} & \gray{\textbf{36.1}} & \gray{***} & \gray{2.8} & \gray{-18.1}    \\\cline{2-13}
& \multirow{2}*{\makecell[l]{DeepSeek-Coder - Base\\\cite{deepseekcoder}}}   & \multirow{2}*{1B}  & Float16 - Baseline & 2.57 & 28.4 & ----- & ----- & ----- & 28.8 & ----- & ----- & -----    \\
& &                                                      & \verylightgray{2-bit} & \verylightgray{0.61} & \verylightgray{\textbf{13.9}} & \verylightgray{***} & \verylightgray{7.56} & \verylightgray{-51.1} & \verylightgray{\textbf{6.6}} & \verylightgray{***} & \verylightgray{23.58} & \verylightgray{-77.1}    \\
& &                                                      & \gray{2-bit + Finetuning} & \gray{0.61} & \gray{\textbf{21.7}} & \gray{***} & \gray{2.41} & \gray{-23.6} & \gray{\textbf{14.7}} & \gray{***} & \gray{6.58} & \gray{-49.0}    \\
\cline{4-13}
& \multirow{2}*{~}   & \multirow{2}*{7B}  & Float16 - Baseline & 13.48 & 45.8 & ----- & ----- & ----- & 41.4 & ----- & ----- & -----    \\
& &                                                      & \verylightgray{2-bit} & \verylightgray{2.27} & \verylightgray{\textbf{35.7}} & \verylightgray{***} & \verylightgray{3.4} & \verylightgray{-22.1} & \verylightgray{\textbf{27.4}} & \verylightgray{***} & \verylightgray{6.21} & \verylightgray{-33.8}    \\
& &                                                      & \gray{2-bit + Finetuning} & \gray{2.27} & \gray{\textbf{36.4}} & \gray{***} & \gray{3.4} & \gray{-20.5} & \gray{\textbf{32.8}} & \gray{***} & \gray{2.84} & \gray{-20.8}    \\
\cline{4-13}
& \multirow{2}*{~}   & \multirow{2}*{33B}  & Float16 - Baseline & 62.16 & 52.1 & ----- & ----- & ----- & 47.3 & ----- & ----- & -----    \\
& &                                                      & \verylightgray{2-bit} & \verylightgray{9.38} & \verylightgray{\textbf{43.4}} & \verylightgray{***} & \verylightgray{2.6} & \verylightgray{-16.7} & \verylightgray{\textbf{34.5}} & \verylightgray{***} & \verylightgray{3.43} & \verylightgray{-27.1}    \\
& &                                                      & \gray{2-bit + Finetuning} & \gray{9.38} & \gray{\textbf{43.0}} & \gray{***} & \gray{3.25} & \gray{-17.5} & \gray{\textbf{38.7}} & \gray{***} & \gray{2.34} & \gray{-18.2}    \\
\bottomrule
\multirow{18}{*}{\rotatebox[origin=c]{90}{\textbf{McEval}}} & \multirow{2}*{\makecell[l]{CodeLlama - Base~\\\cite{roziere2023code}}}   & \multirow{2}*{7B}  & Float16 - Baseline & 13.48 & 12.9 & ----- & ----- & ----- & 29.3 & ----- & ----- & -----    \\
& &                                                      & \verylightgray{2-bit} & \verylightgray{2.26} & \verylightgray{\textbf{11.1}} & \verylightgray{} & \verylightgray{1.3} & \verylightgray{-14.0} & \verylightgray{\textbf{12.8}} & \verylightgray{***} & \verylightgray{8.29} & \verylightgray{-56.3}    \\
& &                                                      & \gray{2-bit + Finetuning} & \gray{2.26} & \gray{13.0} & \gray{} & \gray{0.98} & \gray{0.8} & \gray{\textbf{18.3}} & \gray{***} & \gray{4.08} & \gray{-37.5}    \\
\cline{4-13}
& \multirow{2}*{~}   & \multirow{2}*{13B}  & Float16 - Baseline & 24.25 & 18.9 & ----- & ----- & ----- & 40.9 & ----- & ----- & -----    \\
& &                                                      & \verylightgray{2-bit} & \verylightgray{3.98} & \verylightgray{\textbf{9.4}} & \verylightgray{***} & \verylightgray{6.71} & \verylightgray{-50.3} & \verylightgray{\textbf{22.3}} & \verylightgray{***} & \verylightgray{5.6} & \verylightgray{-45.5}    \\
& &                                                      & \gray{2-bit + Finetuning} & \gray{3.98} & \gray{\textbf{10.4}} & \gray{***} & \gray{4.43} & \gray{-45.0} & \gray{\textbf{27.8}} & \gray{***} & \gray{3.4} & \gray{-32.0}    \\
\cline{4-13}
& \multirow{2}*{~}   & \multirow{2}*{34B}  & Float16 - Baseline & 62.74 & 29.0 & ----- & ----- & ----- & 39.2 & ----- & ----- & -----    \\
& &                                                      & \verylightgray{2-bit} & \verylightgray{9.54} & \verylightgray{\textbf{17.6}} & \verylightgray{***} & \verylightgray{4.43} & \verylightgray{-39.3} & \verylightgray{\textbf{25.2}} & \verylightgray{***} & \verylightgray{3.6} & \verylightgray{-35.7}    \\
& &                                                      & \gray{2-bit + Finetuning} & \gray{9.54} & \gray{\textbf{19.0}} & \gray{***} & \gray{3.27} & \gray{-34.5} & \gray{\textbf{31.6}} & \gray{***} & \gray{1.88} & \gray{-19.4}    \\\cline{2-13}
& \multirow{2}*{\makecell[l]{DeepSeek-Coder - Base\\\cite{deepseekcoder}}}   & \multirow{2}*{1B}  & Float16 - Baseline & 2.57 & 23.8 & ----- & ----- & ----- & 42.0 & ----- & ----- & -----    \\
& &                                                      & \verylightgray{2-bit} & \verylightgray{0.61} & \verylightgray{\textbf{4.4}} & \verylightgray{***} & \verylightgray{28.17} & \verylightgray{-81.5} & \verylightgray{\textbf{8.5}} & \verylightgray{***} & \verylightgray{45.37} & \verylightgray{-79.8}    \\
& &                                                      & \gray{2-bit + Finetuning} & \gray{0.61} & \gray{\textbf{6.9}} & \gray{***} & \gray{12.83} & \gray{-71.0} & \gray{\textbf{15.5}} & \gray{***} & \gray{24.42} & \gray{-63.1}    \\
\cline{4-13}
& \multirow{2}*{~}   & \multirow{2}*{7B}  & Float16 - Baseline & 13.48 & 41.8 & ----- & ----- & ----- & 42.6 & ----- & ----- & -----    \\
& &                                                      & \verylightgray{2-bit} & \verylightgray{2.27} & \verylightgray{\textbf{26.2}} & \verylightgray{***} & \verylightgray{5.68} & \verylightgray{-37.3} & \verylightgray{\textbf{29.1}} & \verylightgray{***} & \verylightgray{2.16} & \verylightgray{-31.7}    \\
& &                                                      & \gray{2-bit + Finetuning} & \gray{2.27} & \gray{\textbf{30.1}} & \gray{***} & \gray{4.5} & \gray{-28.0} & \gray{\textbf{31.0}} & \gray{***} & \gray{2.32} & \gray{-27.2}    \\
\cline{4-13}
& \multirow{2}*{~}   & \multirow{2}*{33B}  & Float16 - Baseline & 62.16 & 55.5 & ----- & ----- & ----- & 57.0 & ----- & ----- & -----    \\
& &                                                      & \verylightgray{2-bit} & \verylightgray{9.38} & \verylightgray{\textbf{36.9}} & \verylightgray{***} & \verylightgray{4.8} & \verylightgray{-33.5} & \verylightgray{\textbf{39.2}} & \verylightgray{***} & \verylightgray{3.58} & \verylightgray{-31.2}    \\
& &                                                      & \gray{2-bit + Finetuning} & \gray{9.38} & \gray{\textbf{39.8}} & \gray{***} & \gray{3.4} & \gray{-28.3} & \gray{\textbf{44.0}} & \gray{***} & \gray{3.38} & \gray{-22.8}    \\
\bottomrule
\end{tabular}
\label{tab:results_rq3}
}
\vspace{-0.2cm}
\end{table*}

As can be seen in the table, an increase in the models' parameters results in a lower performance degradation at the extreme 2-bit quantization level. 
This finding suggests that extreme quantization is less problematic on very large models. However, \tabref{tab:results_rq3} still highlights a statistically significant difference in performance when comparing the 2-bit quantized models to the \texttt{fp16}-precision baselines. The only exception to this trend is CodeLlama 7B on the Python McEval benchmark, where it closes the gap with the base model due to the significant contribution of the mixed calibration.

More interestingly, 2-bit quantizations of large models may outperform their smaller variants at full precision. As an example, CodeLlama 34B quantized to 2 bits per parameter outperforms its smaller 7B-parameter variant at \texttt{fp16} precision on both Python (37.1 \emph{vs} 29.8) and Java (32.7 \emph{vs} 32.2) MultiPL-E benchmarks while requiring less memory ($\sim$9.38GB vs $\sim$13.48GB). However, there is no clear general trend behind this finding, which seems to be model and language-dependent.

For what concerns the fine-tuned variants, we observe that training after quantization can considerably help smaller models to achieve better accuracy. For example, when fine-tuning DeepSeek-Coder 1B, the performance decrease on Java (MultiPL-E) passes from -77.1\% to -49\%, while the one on Python goes from -51.1\% to -23.6\%. We observe that fine-tuning helps reduce the quantization error on smaller models. However, as the number of model parameters increases, we notice that the impact of fine-tuning gradually diminishes and eventually becomes counterproductive. 

\begin{tcolorbox}[top=0pt,bottom=0pt,left=1pt,right=1pt,title=\faLightbulbO~Findings]
	\small
	When the target model features a large number of parameters, 2-bit quantization has a lower negative impact on code generation performance. Given the observed trend, we may expect extremely large code LLMs having hundreds of billions of parameters to possibly be able to support a 2-bit quantization without any significant loss in performance. However, further empirical analyses are needed to generalize this hypothesis.
	\normalsize
\end{tcolorbox}

% !TEX root = ../main.tex
\section{Implications of our Findings}
\label{sec:implications}

The results of our investigation contribute to the growing body of knowledge on quantization methods for code-LLMs, building solid foundations for future research in this area. In particular, our findings empirically show the feasibility of applying advanced quantization techniques, such as AQLM, to reduce the memory footprint of LLMs powering AI systems for SE while maximizing performance. Based on the results of our investigation, we can emphasize two key takeaways:

\noindent \textbf{Code-LLMs can be safely quantized down to 4-bit precision without compromising performances in source code generation tasks}. We empirically showed that quantizing code-LLMs down to 4-bit precision can effectively reduce memory footprint by approximately threefold (x3) without sacrificing performance. This achievement is relevant in different contexts and domains. First, it enables the deployment of larger code models in hardware-constrained environments characterized by limited resources. The consequence is a reduction in energy consumption and an enhancement in terms of scalability, sustainability, and accessibility of advanced AI systems grounded on LLMs for code. Moving along these lines, the use of quantized methods also holds potential in the areas of security and privacy. Outsourcing code to external tools and services to leverage code-LLMs' automation capabilities can pose significant security risks, including the possibility of data leakage \cite{abs-2401-07348,wu2024unveiling,huang2023security}. In contrast, reduced models can be run locally thanks to quantization without significantly compromising performance, eliminating the need to depend on third-party solutions and addressing security and privacy concerns. Moreover, for small businesses and startups that lack access to large-scale computing infrastructure, quantization provides an opportunity for innovation and offers a competitive edge, as these organizations can adopt state-of-the-art AI solutions for automating software engineering-related practices \cite{hou2023large}.
Finally, in the emergent area of agentic AI systems, where multiple smaller models can outperform single monolithic LLMs \cite{belcak2025small}, quantization can effectively enable the deployment of large code models as part of such systems, even in resource-constrained environments.

\noindent \textbf{Additional research is needed on more extreme quantization levels}. While quantization has proven to be an effective tool when bit precision is not pushed to the extreme (\ie 2-bit), we observed that in cases of ultra-low-level quantization, external factors such as the calibration dataset play a significantly more critical role. Therefore, the influence of the calibration dataset is not negligible and requires careful consideration. Optimizing the structure and composition of the calibration dataset becomes essential in such scenarios, as it can greatly affect the overall performance of the quantized model. 
Additionally, with the emergence of \textbf{very} Large Language Models, models with 100 billion parameters or more are becoming the standard rather than the exception, even in software engineering. Codex \cite{humaneval}, the model behind GitHub Copilot, is based on GPT-3, which boasts over 100 billion parameters. However, the application of quantization to models of this size remains largely unexplored. Investigating quantization at this scale offers significant opportunities and could yield new insights that are not yet fully understood, given the current state of affairs.

\section{Threats to Validity} \label{sec:threats}

\textbf{Construct validity.} This threat is mainly related to the measurements made to address our research questions. As far as models' accuracy is concerned, we leverage the \textit{pass@k}  measure, used in previous work \cite{humaneval,cassano:tse2023,cassano2023knowledge,wei2023towards} including the Wei \etal study we aimed at partially replicating~\cite{wei2023towards}.

\textbf{Internal validity.} These threats mainly concern the choice of hyperparameters for the quantization process, for the end-to-end fine-tuning, and for the inference phase. In \secref{sec:design}, we explain, for the different phases of our study, how the different hyperparameters have been chosen. More specifically, in some cases, we based our choices on what was done in previous work, including the original AQLM paper \cite{egiazarian2024extreme} and the quantization study we replicated \cite{wei2023towards}. In other cases, we made empirical choices, for example when determining the number of epochs in the end-to-end fine-tuning.

\textbf{Conclusion validity.} To support our findings, we leverage suitable statistical procedures, \ie McNemar's test (adjusted \emph{p}-values via Benjamini-Hochberg procedure \cite{yoav:jstor1995}) and Odds Ratio.

\textbf{External validity.}
Concerning the studied LLMs, we focused on two widely used models of code, CodeLLama and DeepSeek-Coder. Also, we experimented with different model sizes, up to 34B. However, it is possible that other code-based LLMs or general-purpose LLMs may be differently impacted by the quantization process. Although we did not experiment with larger models (\eg CodeLlama 70B), the trend observed in RQ$_3$ suggests that the benefits of extreme quantization are likely to extend to $>$34B models as well.
Another threat to external validity concerns the generalizability to other tasks. In this work, we evaluated the effects of quantization on code generation tasks. Therefore, the obtained results may not generalize to other SE-related tasks.

% !TEX root = main.tex
\section{Conclusion and Future Work} 
\label{sec:conclusion}

Previous work \cite{wei2023towards} showed quantization to be an effective way of managing the size of state-of-the-art LLMs that power AI-driven solutions for the software engineering domain. However, the recent surge in quantization techniques and code-LLMs made us question on what would be the pros and cons of quantization given (i) new advancements, including not only adopting extreme quantization precision levels but also different types of calibration datasets and (ii) large code models featuring up to 34 Billion parameters.

To further explore quantization for code-LLMs, we partially replicated the foundational study by Wei \etal, confirming that quantization, as proposed, is an effective approach for reducing memory usage. Building on their work, we expanded our study's scope by focusing on source code generation, incorporating two programming languages—Java and Python—and evaluating two state-of-the-art code models: (i) CodeLlama and (ii) DeepSeek-Coder.

We investigated low-bit quantization (\ie 3 and 2 bits), pushing the boundaries of extreme quantization for code models while also incorporating different calibration datasets: one consisting of randomly selected elements, another featuring a mix of code and technical natural language, and a third made up entirely of code. 

Our findings revealed that code models can be quantized down to 4 bits, achieving a 70\% reduction in memory usage while maintaining the original performance of the non-quantized model. Moreover, we observed that the choice of calibration dataset significantly impacts the model's performance. Specifically, when applying low-bit quantization ($\leq$4 bits), the selection of curated examples---instead of randomly sampled ones---was beneficial in mitigating performance degradation. Last but not least, we found that larger code models (with $\geq$33 billion parameters) were more resilient to information loss during quantization, showing a smaller drop in performance compared to smaller models when the bit precision was reduced to the point of ultra low-bit quantization, \ie 2 and 3 bits.

In conclusion, our research not only reaffirms the benefits of quantization, as highlighted by Wei \etal, but also extends the possibilities of what can be achieved with modern code models when using state-of-the-art quantization techniques. 

Our next step is to assess the impact of quantization on a wider range of code-related tasks, such as code summarization and program repair, beyond just code generation. We also plan to investigate the effects of quantization on other dimensions, such as inference latency and energy consumption.
These additional analyses would offer a more comprehensive understanding of how quantization affects code models across the entire spectrum of code-related tasks, thereby helping to generalize our findings further.

% !TEX root = main.tex
%%%%%%%%%%%%%%%%%%%%%%%%%%%%%%%%%%%%%%%%
%%%%%%%%%%%%%%%%%%%%%%%%%%%%%%%%%%%%%%%%
\section{Acknowledgments}\label{sec:acknowledgments}
%%%%%%%%%%%%%%%%%%%%%%%%%%%%%%%%%%%%%%%%
%%%%%%%%%%%%%%%%%%%%%%%%%%%%%%%%%%%%%%%%
Di Penta acknowledges the Italian ‘‘PRIN 2022’’ project TRex-SE: ‘‘Trustworthy Recommenders for Software Engineers’’, grant n. \\ 2022LKJWHC. Dr. Mastropaolo acknowledges the support of the National Science Foundation (NSF) under grant CCF-2451058. \\
Giagnorio and Bavota thank the Swiss National Science Foundation (SNSF) for the funding provided under the project ``PARSED'' (grant agreement No. 219294). Giagnorio also thanks CHOOSE for sponsoring his trip to the conference.

\bibliographystyle{ACM-Reference-Format}
\bibliography{main}

\end{document}